# EXPLORING THE NONLINEAR CLOUD AND RAIN EQUATION

Ilan Koren, Eli Tziperman and Graham Feingold

Department of Earth and Planetary Sciences, Weizmann Institute, Rehovot 76100, Israel

Department of Earth and Planetary Sciences, and School of Engineering and Applied Sciences, Harvard University, Cambridge, Massachusetts, 02138 USA

Chemical Sciences Division, NOAA Earth System Research Laboratory, Boulder, Colorado, 80305 USA,

Corresponding author: Ilan Koren, e-mail: ilan.koren@weizmann.ac.il

## Abstract

Marine stratocumulus cloud decks are regarded as the reflectors of the climate system, returning back to space a significant part of the income solar radiation, thus cooling the atmosphere. Such clouds can exist in two stable modes, open and closed cells, for a wide range of environmental conditions. This emergent behavior of the system, and its sensitivity to aerosol and environmental properties, is captured by a set of nonlinear equations. Here, using linear stability analysis, we express the transition from steady to a limit-cycle state analytically, showing how it depends on the model parameters. We show that the control of the droplet concentration ($N$) the environmental carrying-capacity ($H_0$) and the cloud recovery parameter ($\tau$) can be linked by a single nondimensional parameter $\left(\mu = \sqrt{N}/(\alpha\tau H_0)\right)$, suggesting that for deeper clouds the transition from open (oscillating) to closed (stable fixed point) cells will occur for higher droplet concentration (i.e. higher aerosol loading).

The analytical calculations of the possible states, and how they are affected by changes in aerosol and the environmental variables, provide an enhanced understanding of the complex interactions of clouds and rain.

**We describe and explore a delay differential equation that captures key elements of the interplay between cloud formation and depletion by rain, and how it is regulated by atmospheric aerosol. We analytically obtain the Hopf bifurcation points that describe transitions between a stable fixed point, which implies a balance between cloud formation and depletion (a steady state), to a limit cycle behavior that is related to cycles of formation of thicker clouds that are later consumed by stronger rain. At distances in the parameter space further away from the bifurcation point, the system exhibits a period doubling route to chaos. Exploring how the model transitions depend on the environmental conditions and aerosol concentration sheds new light on the cloud's sensitivity to the interplay between key parameters in nature, and specifically to possible anthropogenic aerosol effects on cloud properties and transitions between cloud states.**

## Introduction

Marine stratocumulus cloud decks forming over dark, subtropical oceans are regarded as the reflectors of the atmosphere.[1] The decks of low clouds 1000s of km in scale reflect back to space a significant portion of the direct solar radiation and therefore dramatically increase the local albedo of areas otherwise characterized by dark oceans below.[2,3] This cloud system has been shown to have two stable states: open and closed cells. Closed cell cloud systems have high cloud fraction and are usually shallower, while open cells have low cloud fraction and form thicker clouds mostly over the convective cell walls and therefore have a smaller domain average albedo.[4-6] Closed cells tend to be associated with the eastern part of the subtropical oceans, forming over cold water (upwelling areas) and within a low, stable atmospheric marine boundary layer (MBL), while open cells tend to form over warmer water with a deeper MBL. Nevertheless, both states can coexist for a wide range of environmental conditions.[5,7] Aerosols, liquid or solid particles suspended in the atmosphere, serve as Cloud Condensation Nuclei (CCN) and therefore affect the concentration of activated cloud droplets.[8] Changes in droplet concentration affect key cloud properties such as the time it takes for the onset of significant collision and coalescence between droplets, a process critical for rain formation. The onset of significant collision-coalescence process can thus be represented by a delay factor.[9,10]

The emergent behavior of the coevolution of cloud and rain has been shown to be captured by a set of dynamical equations with a delayed sink term.[11] Numerical analysis of these equations yields bifurcation points that separate different dynamical regimes. The first point marks a shift from a steady-state (stable fixed point) in which the rain consumes the cloud at the exact rate of cloud replenishment, to oscillations (limit cycle) of stronger rain that depletes the cloud that created it and therefore dissipates until the cloud is thick enough to reform rain. These results were shown to provide insights into naturally occurring closed and open cells in marine stratocumulus cloud systems,[12] and the processes underlying transitions between these states. The oscillating branch of the solutions represents open cell clouds that typically produce stronger rain,[1,13,14] and as the rain depletes the cloud water and suppresses the updraft by evaporating below cloud base, the average cloud tends to last for shorter durations.[6] The closed cells tend to produce very little drizzle and their morphology remains stable for more than 10 h despite the fact that their characteristic scale suggests a theoretical lifetime of only ~1hr.[12,15,16] The one-dimensional, time-delay equations for cloud thickness ($H$) and for droplet concentration ($N$) were later coupled by a spatial dynamical feedback to a set of oscillators that produce spatial patterns of cellular convection similar to the ones produced in detailed cloud resolving models and seen in nature.[17]

The nature of the transitions from open to closed cellular convection has been studied using both large eddy simulation (LES) and the cloud and rain equations.[18] This study showed that the transition between closed and open cellular cloud states shows hysteresis as function of the aerosol loading. Such behavior is expected in a Delay-Differential-Equation (DDE) as the solution depends on the past history of the delayed element.

Here we explore the cloud and rain equation response to small perturbations around the steady state, which allows analytical exploration of the nature of the damped oscillations toward the fixed points, and of the first bifurcation point that marks the transition from a fixed point to a limit cycle. For completeness we further explore the transition toward chaotic behavior. This transition occurs in a nonphysical regime in the current simple model, yet it is possible that subsequent studies will find such a transition in a physical regime of somewhat more detailed model. The linear stability analysis allows us to better understand transitions between states and how they depend on changes in aerosol and the environmental variables, which provides new insights into the complex interactions between clouds, aerosol, and rain. As aerosol concentration strongly affects the cloud droplet concentration ($N$), changes in the parameter $N$ imply here changes in the aerosol loading.

## 2. Model Equations

Time delay differential equations (DDE) are used in many dynamical systems,[19] including population dynamics,[20] neural networks,[21] El Nino,[22] and more. As even a simple linear DDE requires an infinite number of initial conditions to initialize the delay term, it is formally equivalent to a PDE or to an infinite system of ordinary differential equations in terms of its number of degrees of freedom. DDEs can display complex behavior, some of them chaotic, and therefore often cannot be solved analytically.[19]

Using the notation $H(t-D)$ for the $H$ value in $D$ time units before the time $t$, the cloud and rain equation can be coupled to an aerosol equation as follows,[11]

$$\frac{dH}{dt} = \frac{H_0 - H}{\tau} - \frac{\alpha}{\sqrt{N}} H^2(t-D)$$

$$\frac{dN}{dt} = \frac{N_0 - N}{\tau_2} - cR(t-D)N(t-D)$$

where the time dependent variables are the cloud depth (*H*), and the droplet concentration (*N*). $R$ is the rain-rate, which is a function of $H$ and $N$, and $\alpha \approx 100\ [day^{-1}m^{-2.5}]$ is a scaling constant that links cloud depth, droplet concentration and rain rate. The $\alpha$ value was determined both theoretically and from measurements.[23-25] The environmental and droplet (aerosol) conditions are represented by $H_0$ and $N_0$ as the cloud-depth and droplet concentration carrying capacities, $\tau$ and $\tau_2$ are characteristic times for reaching the carrying capacity values under no sink conditions, $D$ is the delays that represents past states that control the current sinks. Note that following theoretical and modeling studies the delayed rain sink term depends on the inverse of the square root of *N*.[26]

Here we consider the basic cloud and rain equation in which *N* is a free parameter,

$$\frac{dH}{dt} = \frac{H_0 - H}{\tau} - \frac{\alpha}{\sqrt{N}} H^2(t-D). \qquad (1)$$

Such a representation of the problem assumes that changes in the aerosol concentration are relatively small. It represents cases for which the source of aerosols is steady or when the aerosol consumption by drizzle is relatively small. This reduces the problem to a first order nonlinear DDE controlled by four parameters: (i) $H_0$ - the cloud carrying capacity parameter that represents the systems maximal potential for cloud depth. (ii) $\tau$ - the characteristic cloud recovery time. (iii) *N*, which controls the strength of the sink term (rain), and (iv) $D$ - the time delay that represents the time it takes to convert cloud water into rain by stochastic microphysical collection processes.[11]

Equation 1 can be nondimensionalized by replacing $H$ with the normalized height $h = H/H_0$ and $t$ with $t^* = t/\tau$. When translating the equation for $dH/dt$ to $dh/dt^*$, model parameters *N*, $H_0$ and $\tau$ are replaced by a single parameter,

$$\mu = \frac{\sqrt{N}}{\alpha \tau H_0} \qquad (2)$$

and Eq. 1 is therefore transformed to a simpler nondimensional form,

$$\frac{dh}{dt^*} = 1 - h - \frac{1}{\mu} h^2(t^* - D^*), \qquad (3)$$

where $D^* = D/\tau$ is the nondimensional delay. Throughout this paper we will refer to the nondimensional version of the problem (Eq. 3) as the Cloud and Rain (C&R) equation. When solving Eq. 3 for the steady state case ($h_{sts}$), for which the derivative ($\frac{dh}{dt^*}$) vanishes, $h = h_{sts}$ is constant in time and the delay does not play a role in determining $h_{sts}$. In such a case Eq. 3 reduces to a simple one-parameter polynomial whose physical solution (allowing only positive $h$ values) is

$$h_{sts} = \sqrt{\frac{\mu^2}{4} + \mu} - \frac{\mu}{2}. \quad (4)$$

The fact that $N^{1/2}$ and the cloud depth carrying capacity appear only via their ratio in a single nondimensional parameter ($\mu$) has an important implication that will be discussed later. We note the $h_{sts}$ is a fixed point that can be either stable (to which the C&R equation converges in the steady-state) or non-stable. For all cases, if the equation's initial conditions are equal to $h_{sts}$ throughout the delay period, the solution will remain $h_{sts}$ for all later times. Next we will use linearized stability analysis to explore the system's response to small perturbations from $h_{sts}$.

## 3. Analytical solution for the case of no delay

When the delay is set to zero the C&R equation becomes a first order, nonlinear ordinary differential equation with a quadratic term (a Riccati equation)[27] that can be transformed to a second order linear ordinary differential equation and has an analytical solution of the form,

$$h = \frac{\mu}{2}\left[-\varsigma \tanh\left(\frac{k_1}{2}\varsigma - \frac{t^*}{2}\varsigma\right) - 1\right], \quad (5)$$

where $k_1$ is an integration constant and $\varsigma = \sqrt{\frac{4}{\mu} + 1}$ is a positive nondimensional number. Prescribing the initial conditions $h(t = 0) = 0$, to show how the cloud develops in time, yields $k_1 = \frac{2}{\varsigma}\tanh^{-1}(-1/\varsigma)$. The hyperbolic tangent function approaches 1 for positive argument, or -1 for negative argument, with a sharp transition near zero. For large enough $t^*$ the hyperbolic tangent component can therefore be replaced with $-1$ such that the solution for $h$ converges to the steady state solution of Eq. 4. The transition in time from: $h = 0$ to $h = h_{sts}$ is smooth and monotonic with no oscillations, as for an over-damped oscillator. This behavior will be linked to the stability analysis of the fuller equation with delay in the next section.

## 4. Stability analysis around $h_{sts}$

As stated earlier, $h_{sts}$ is a fixed point for all parameter values. We can therefore perform a stability analysis of the C&R equation to investigate the response to small perturbations around the fixed point for different values of the model parameters.

Let $\delta$ be a small perturbation around the fixed point $h_{sts}$ such that $h = h_{sts} + \delta(t)$, thus $\frac{d\delta}{dt^*} = \frac{dh}{dt^*}$. Linearizing around $h_{sts}$ and neglecting terms nonlinear in $\delta$ yields:

$$\frac{d\delta}{dt^*} = -\delta - \frac{2}{\mu} h_{sts}\delta(t^* - D^*). \tag{6}$$

Expressing δ as an exponent $\delta = e^{\beta t^*}$ allows separation of the contribution of the delay, and transforms Eq. 6 to

$$\beta e^{\beta t^*} = -e^{\beta t^*} - \frac{2}{\mu} h_{sts} e^{\beta t^*} e^{-\beta D^*}, \tag{7}$$

thus yielding a transcendental equation for $\beta$:

$$\beta = -1 - \frac{2}{\mu} h_{sts} e^{-\beta D^*} \tag{8}$$

The exponent $\beta$ may, in general, be complex. Its real part, $Re\{\beta\}$, determines the stability of the perturbations, and its imaginary part, $Im\{\beta\}$, the frequency as the solution converges or diverges from the fixed point. For $Re\{\beta\} < 0$ & $Im\{\beta\} = 0$ the system is in a state similar to an overdamped oscillator converging to the fixed point with no oscillations (a continuation of the case of no delay). The case $Re\{\beta\} < 0$ & $Im\{\beta\} > 0$ describes decaying oscillations toward the steady state. $Re\{\beta\} > 0$ generates an unstable fixed point where the perturbation is amplified, leading to steady oscillations. Points in the parameter space for which $Re\{\beta\} = 0$ describe the first bifurcation in which the system transforms from having a stable to a non-stable fixed point.[28]

The transcendental equation for $\beta$ (Eq. 8) has a closed form solution based on the Lambert $W$ function that solves $W(z)e^{W(z)} = z$, which is often used for DDE analysis,[29]

$$\beta = \frac{1}{D^*} W\left(-2\frac{h_{sts}}{\mu} D^* e^{D^*}\right) - 1. \tag{9}$$

Inserting the expression for $h_{sts}$ from Eq.4, into the argument of the Lambert $W$ function in Eq. 9 ($\xi$) yields:

$$\xi = -2\frac{h_{sts}}{\mu} D^* e^{D^*} = \left(1 - \sqrt{1 + \frac{4}{\mu}}\right) D^* e^{D^*}. \tag{10}$$

For the physical parameter range the argument $\xi$ is always negative and is composed of two factors, one a function of $\mu$ and one of $D^*$. $\xi$ grows monotonically with $\mu$ (becomes less negative and smaller in absolute value) and decreases with $D^*$. The main branch of the Lambert W function is a real and negative number as long as its argument is real, negative number equal or larger than $-e^{-1}$ (where it reaches its global minimum $W(-e^{-1}) = -1$ ). Therefore, since $\xi$ is always negative, Eq 9. reveals that as long as $\xi \geq -e^{-1}$, $\beta$ will have a real and negative value. For such cases small perturbation $\delta = e^{\beta t^*}$ decays exponentially to the fixed point $h_{sts}$ with no oscillations (an overdamped regime). Inserting the $\xi$ expression (Eq. 10) and solving this condition explicitly, shows that as long as

$$\frac{1 + D^* e^{D^*+1}}{D^* e^{D^*+1}} > \sqrt{\frac{\mu + 4}{\mu}} \tag{11}$$

$\beta$ is real and negative. For a given nonzero value of $\mu$, as $D^*$ approaches zero the left side of Eq. 11 increases rapidly such that the condition is always fulfilled and the solution is similar to the no-delay case of $D^* = 0$, which we showed to be overdamped (Eq. 5). We define a critical delay ($D_c^*$) which satisfies the equality $\frac{1+D^* e^{D^*+1}}{D^* e^{D^*+1}} = \sqrt{\frac{\mu+4}{\mu}}$, for which $\xi = -e^{-1}$ and therefore where $Re\{\beta\}$ reaches its minimal value of,

$$\beta = -\left(\frac{1}{D_c^*} + 1\right) \qquad (12)$$

and where the solution's convergence to the fixed point is fastest.

For $\xi < -e^{-1}$, $Im\{\beta\}$ is not zero. The oscillations decay toward the stable fixed point as long as $Re\{\beta\} < 0$, namely based on Eq. 9, as long as $Re\{W(\xi)\} < D^*$, while the first bifurcation point, i.e., the transition to a limit cycle state, occurs when $Re\{\beta\} = 0$, namely $Re\{W(\xi)\} = D^*$.

Writing the explicit equation for the value of the model parameters at the bifurcation point,

$$Re\left\{W\left(\left(1 - \sqrt{1 + \frac{4}{\mu}}\right) D^* e^{D^*}\right)\right\} = D^* \qquad (13),$$

reveals an equation that resembles one of the definitions of the Lambert W function (i.e. $W(ze^z) = z$) with an additional negative scalar $\psi = \left(1 - \sqrt{1 + \frac{4}{\mu}}\right)$. The argument of the Lambert W function in Eq. 13 is therefore a negative real number. If the solution for $W\left(\psi D^* e^{D^*}\right) = z$, and $z = x + iy$, solving and for $x$ and $y$ while requiring that $Im\{ze^z\} = 0$ yields $x = D^*$ and $\psi = \frac{1}{\cos y}$, which implies that $|\psi| \geq 1$ (for a more detailed proof see the lemma in appendix A). Therefore Eq. 13 can be satisfied only when $\left|1 - \sqrt{1 + \frac{4}{\mu}}\right| \geq 1$, namely when $\mu \leq 4/3$. We define this limit as $\mu_{lim} = 4/3$; The bifurcation points exist only when $\mu < \mu_{lim}$. Therefore, any point for which $\mu \geq \mu_{lim}$ is a stable fixed point.

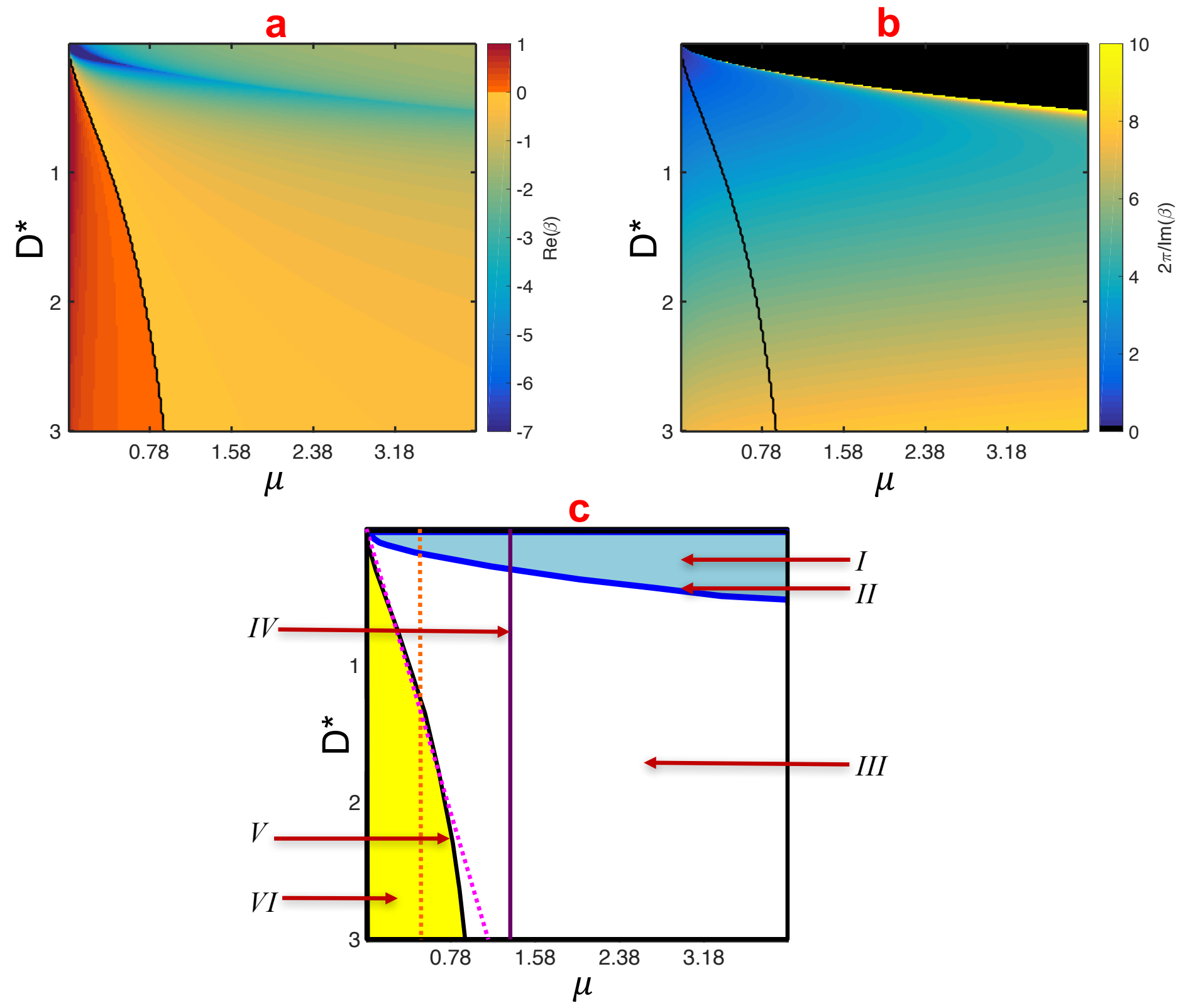

*Figure 1. β, the linear stability parameter as a function of μ and $D^*$. Panel 'a' shows the $Re\{\beta\}$ values, panel 'b' shows the nondimensional oscillation period $2\pi/Im\{\beta\}$, and panel 'c' shows a schematic map of the μ vs. $D^*$ parameter space, marking key features. Six distinct regimes are shown here: (I) The area for which $Im\{\beta\} = 0$ marking the overdamped state (light blue area on panel c). (II) On the edge of the first area, the line for which $Re\{\beta\}$ has a local minimum (shown as narrow blue valley on panel 'a') marking the critical values for which the fastest local response to perturbation converges back to $h_{sts}$ (blue line on panel c). (III) The area in which $Re\{\beta\} < 0$ and $Im\{\beta\} > 0$ marking damped oscillations toward the stable fixed point (white area in panel c). (IV) The $\mu=\mu_{lim}$ (4/3) line (purple vertical line on panel c) marking the limit for non-stable fixed point. All points with $\mu>\mu_{lim}$ will be in stable fixed point state. (V) The black contour that marks the first bifurcation point for which $Re\{\beta\} = 0$ and the system state changes from $h_{sts}$, i.e,, stable fixed point to oscillations. (VI) The area in which $Re\{\beta\} > 0$ is marked between the first bifurcation contour and the Y axis for which $h_{sts}$ is not a stable fixed point. Numerical simulations with selected parameters along the orange (fixed μ) and along the magenta (varying τ, fixed D, $H_0$ and N) dotted lines shown in panel 'c' are discussed below.*

Figure 1 shows $\beta$ as a function of the two controlling parameters: $\mu$ and $D^*$. Five distinct features can be seen in this parameter-space (the $\mu$ vs. $D^*$ space):

I. The area in which $Im\{\beta\} = 0$, and therefore where the system is overdamped, is shaded light blue in panel 'c', revealing that for lower delay values the perturbation decays toward the fixed point ($h_{sts}$) with no oscillations.

II. The line along which $Re\{\beta\}$ has a local minimum is shown in panel 'a' (narrow blue "valley") and marked as a blue contour in panel 'c', marking the edge of the overdamped regime. Points along this line mark the fastest decay of the perturbation toward $h_{sts}$ for a given value of $\mu$. For a given $\mu$, the corresponding delay values along this line are defined as the critical delays ($D_c^*$), which can be calculated by Eq. 12.

III. The area in which $Re\{\beta\} < 0$ & $Im\{\beta\} > 0$ is between the critical delay line and the black contour shown on the left panel (white area in panel 'c'). Values of $D^*$ and μ in this regime will lead to damped oscillations. The shading of the period as function of the two model parameters (Fig 1, panel 'b') reveals a decrease of the period near the critical delay line up to a local minimum followed by a monotonic increase (reduction of the frequency) as a function of the delay.

IV. Solutions corresponding to points in the parameter domain to the right of the vertical $\mu_{lim}$ line (i.e. $\mu > \mu_{lim}$, purple vertical line in panel 'c') can only be of stable fixed point type.

V. The first bifurcation line is marked by the black contour line, for which $Re\{\beta\} = 0$, and therefore $Re\{W(\xi)\} = D^*$. This line shows the transition at which $h_{sts}$ shifts from a stable to a non-stable fixed point. For small $D*$ values the bifurcation point described in Eq. 13 occurs for small $\mu$ values that monotonically grow as $D*$ increases approaching the $\mu$=4/3 limit.

VI. The area in which $Re\{\beta\} > 0$ is left of the black contour (panel 'c'). In this part of the parameter space $h_{sts}$ becomes an unstable fixed point and therefore perturbations from the fixed point will shift the solution to a limit cycle or, as will be shown later, to a period doubling route to chaos.[28]

To further explore the nature of the transition from stable fixed point to non-stable, we run a set of numerical simulations around the first bifurcation point. Because $h_{sts}$ is independent of $D^*$, it is convenient to change it while holding μ constant such that $h_{sts}$ is unchanged (see Eq. 4). Figure 2 shows that the equation undergoes a supercritical Hopf bifurcation, showing that as the delay is increased, the amplitude of the oscillation indeed increases as $\sim\sqrt{D^* - D_0^*}$, away from the bifurcation point ($D_0^*$). For completeness we also present the evolution for the non-physical regime for which the lower $h$ values are negative, and find that the solution undergoes a period-doubling route to chaos, and the ratios of the distance between the values of $D*$ at the period-doubling points converge to the Feigenbaum constant.[30,31]

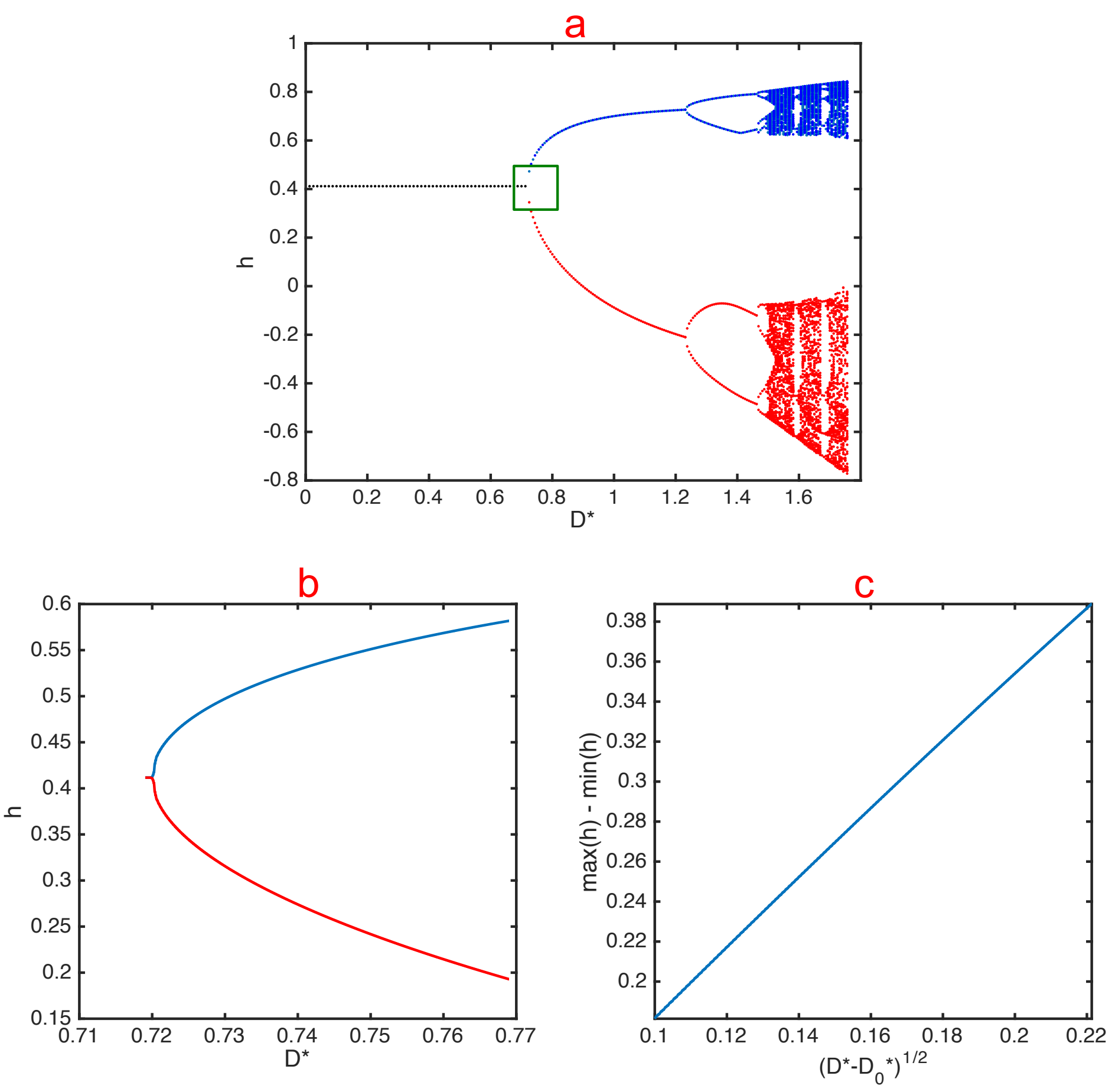

*Figure 2. Numerical simulations of the nondimensional C&R equation (Eq. 3) as a function of D* for a fixed μ=0.29, such that $h_{sts}$ is constant for all runs. Panel 'a' shows the system evolution from a stable fixed point through a limit cycle state when passing the first bifurcation point ($D^*_0$ ~0.72) and later to a period doubling route to chaos. The black dotted line shows the case for which the system is in a stable steady state, ($h_{sts}$=0.41). The blue and red lines mark local maxima and minima of the function. Panel 'b' is a zoom-in around the first bifurcation point (marked in green rectangle in panel 'a'). On panel 'c' the amplitude differences are plotted against the square root of the distance from the bifurcation point $\sqrt{D^* - D^*_0}$. As expected for a supercritical Hopf bifurcation such a graph shows linear relations.*

Recall that we solved here the nondimensional form of the C&R equation for which 3 out of the 4 model parameters were replaced by $\mu$, reducing the dimensionality to only 2 nondimensional parameters. However, the interpretation of the results in the original dimensional parameter space is not always straightforward, because changing a dimensional parameter corresponds to moving along some curve in the 2d $\mu$ vs. $D^*$ nondimensional parameter space. In particular we note that both nondimensional parameters $D*$ and μ are normalized by $\tau^{-1}$ while $\mu$ is also normalized by $\sqrt{N}/H_0$. Varying $\sqrt{N}/H_0$ while keeping other parameters fixed corresponds to a straight horizontal line parallel to the $\mu$ axis in the $\mu$ vs. $D^*$ space, showing that as $\sqrt{N}/H_0$ increases (i.e., increase in $\mu$) the system's state will shift from an unstable to a stable fixed point. Changes in the dimensional delay $D$, corresponds to a straight vertical line in the $\mu$ vs. $D^*$ space, parallel to the $D^*$ axis (e.g. the dotted orange vertical line on Fig 1, panel 'c') for which $\mu$ is held constant. An increase in $D$ leads to a transition from stable to non-stable fixed points. As shown in Eq. 13, the $D^*$ values for which the Hopf bifurcation point occurs depend on $\mu$, and for $\mu > \mu_{lim}$ the equation will have a stable fixed point for any $D^*$. Varying $\tau$ while holding all other variables fixed is described by an inclined line in the $\mu$ vs. $D^*$ space, approaching the origin (when $\tau \to \infty$) for which the slope depends on $\sqrt{N}/H_0$. As illustrated in the dotted magenta line on Fig 1 (panel 'c'), some of these lines can cross the Hopf bifurcation contour twice, showing a transition from stable fixed point to non-stable and then to stable again, as $\tau$ increases.

To illustrate how $\beta$ affects the solution, Fig 3 shows simulations for parameters corresponding to two lines in the $\mu$ vs. $D^*$ parameter space: first (dotted orange line in Fig, 1 panel 'c'), when $D$ is the free variable and all other parameters are fixed (similarly to the runs in Fig. 2 for which $H_0$ = 1000 m, $N$ = 16 cm$^{-3}$, $\tau$ = 20 min, therefore $\mu$ = 0.29, yielding a constant fixed point), and second, when $\tau$ is the free variable ($H_0$ = 1000 m, $N$ = 16 cm$^{-3}$and $D$ = 15 min, marked by the dotted magenta line in Fig 1, panel 'c'). The two numerical simulations were run for five values of $\beta$ each (marked on Fig. 3, upper panels) controlled by changes in $D$ (left column) and $\tau$ (right column), demonstrating qualitatively different solutions for $\beta$ and therefore different stability regimes of the dynamics around the steady state. Each simulation was initiated with the fixed-point cloud height, $H = H_{sts}$, and the thickness therefore remained constant in time, until a small perturbation was introduced. As shown in Eq. 4 $h_{sts}$ is a function of $\mu$ only; hence for the cases in which the delay is the free parameter, $h_{sts}$ remains the same for all simulations (Fig. 3 left column). For the first 3 delay varying simulations (Fig 3. Panels 'b' 'c' and 'd') $Re\{\beta\} < 0$ and therefore the perturbations decay back to the $H_{sts}$ which is a stable fixed point. In the case of panel 'b' $Im\{\beta\} = 0$ and therefore the system is in the overdamped regime and solutions decay to $H_{sts}$ with no oscillations. As the values of $\beta$ are closer to the bifurcation point the decay to the fixed point is slower. On the same note when $Re\{\beta\} > 0$ (panels 'e' and 'f') the closer the $\beta$ values are to the bifurcation point, the slower is the shift to a steady limit cycle state. The $\tau$ varying simulations (Fig 3. right column) show that the system can enter and exit the non-stable fixed point regime crossing the Hopf bifurcation contour twice, and therefore will have a stable fixed point for relatively small $\tau$ values (panels 'h' and 'i') and for relatively high $\tau$ values (panel 'l') and between them be in the limit cycle state (panels 'j' and 'k').

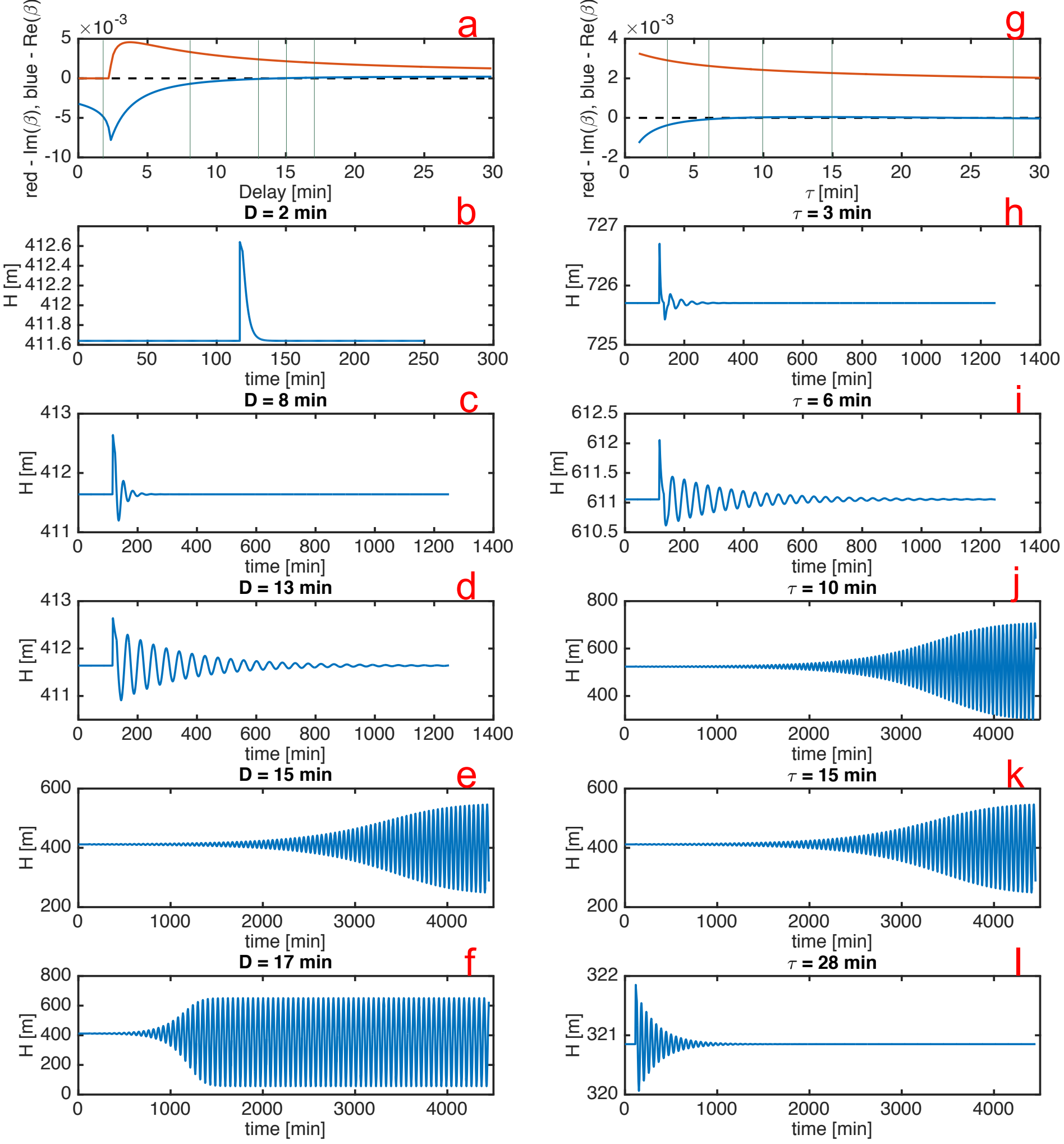

*Figure 3. Illustrations of the system response to small perturbations around the fixed point along two lines in the μ vs. D* parameter space. The left column is for fixed μ that yields fixed steady state height ($h_{sts} = \sim 412\ m$) described as the dotted vertical orange line in figure 1 (panel 'c') and the right column is for varying τ (fixed D, $H_0$ and N)) described as the inclined dotted magenta line crossing the origin in figure 1 'c'. The upper panels show the real and imaginary parts of β as a function of D (left) and τ (right). Five distinct β values (marked in each panel) describe different states of the numerically simulated system. In the left column: Panel 'b' is in the overdamped regime for which Im(β)=0 and Re(β)<0. The conversion toward the fixed point is fast with no overshooting (oscillations). Panels 'c' and 'd' are in the damped oscillatory regime for which Im(β)>0 and Re(β)<0 showing the system response to the perturbation with damped oscillations. Re(β) is more negative for case 'c' and therefore the damping factor of the exponent β is larger and the system converges to the fixed point faster. Re(β)>0 for the cases shown in panels 'e' and 'f' indicate that the fixed point is not stable and that the system will shift to a limit-cycle state. Re(β) is positive and larger for case 'f' and therefore the system deviates faster from the fixed point to the oscillating state. The right column shows that when τ is the free parameter, the system can enter and exit the non-stable fixed point regime crossing the Hopf bifurcation contour twice: In panel 'h' $\mu > \mu_{lim}$ and therefore the fixed point is stable. In panel 'i', $\mu < \mu_{lim}$ but still Re{β} <0 and therefore the oscillations decay toward the stable fixed point. In panels 'j' and 'k' Re{β} > 0 indicating that the fixed point is not stable and the system is in a limit-cycle state while in panel 'l' D*~0.5 and the system shifts again to the steady-state area of the μ vs. D* plane near the origin.*

## 6. Discussion

In this paper we consider warm marine stratocumulus clouds for which all hydrometeors are liquid (cloud droplets and raindrops). Cloud droplets nucleate on aerosol particles that serve as Cloud Condensation Nuclei (CCN). Raindrops form when larger droplets collide and collect the smaller ones. The efficiency of transferring cloud droplets to rain correlates positively with the first two moments of the droplet size distribution, i.e. the average size and the variance. An increase in the aerosol concentration increases the concentration of activated cloud droplets. This implies that more droplets are competing for the available vapor and therefore their average size will decrease.[8,32] Such changes incur a suppression of drop collection, so that variance decreases. The result is a delay in the onset of rain formation, and for marine stratocumulus a decrease in the rain-rate up to a point of complete rain suppression.[9,33]

We have analyzed the cloud and rain (C&R) equation for which the droplet (aerosol) concentration is assumed fixed and is prescribed. Such an assumption can be justified when the source of aerosols is steady (either local or due to long range transport) and when the aerosol consumption by drizzle is relatively small. In such cases, fixed $N$ (often also assumed in cloud resolving models) allows one to study the coevolution of cloud and rain in a model of reduced complexity.

The C&R question has four parameters ($\tau$, $D$, $H_0$ and $N$). When nondimensionalizing the equation, the environmental parameters $\tau$, and $H_0$ and the droplet concentration $N$ are replaced by a single parameter $\mu = \sqrt{N}/(\alpha\tau H_0)$. Apart from reducing the complexity, this offers important physical insights into the interplay between the environmental and aerosol properties with respect to the system's stability.

We performed linear stability analysis which yielded an analytical expression for the stability as a function of the two parameters ($\mu$ and $D^*$). The C&R equation forms a very rich solution space, ranging from a stable over-damped convergence trough fixed point, to a limit cycle, to chaotic behavior. Each solution type occupies a distinct regime in the $\mu$ vs. D* parameter space.

The nondimensional delay $D^*$ and $\mu$ have opposite effects on the stability. Larger $\mu$ values (larger aerosol concentration, all other variables being equal, hence smaller sink term) imply a more stable solution, while an increase in $D^*$ leads to an instability of the steady solution. In the limit we have shown that any point for which $\mu > \mu_{lim}$ yields solutions that are characterized by a stable fixed point.

For a small $D^*$ and a large $\mu$, perturbations decay back to the steady solution with no oscillations. For larger $D^*$ the system shifts, through a critical delay $D_c^*$, to a damped oscillatory regime, and later, crossing the first bifurcation point, to a limit cycle regime, and finally to a chaotic regime.

As expected, an increase in the aerosol and therefore in the droplet concentration *N,* decreases the rain-rate. Reduction in the rain-rate can shift the system state from a limit cycle to coexistence of steady (weak) rain and fixed cloud thickness (stable fixed point solution). The $\mu$ parameter is linear in $\sqrt{N}/H_0$ and does not depend on $D*$. This suggests that clouds forming in atmospheric conditions that allow thicker clouds will shift from a limit cycle state to a steady-state under higher aerosol concentrations, and therefore that the shift from open to closed cells would occur at higher aerosol concentrations for a thicker marine boundary layer.

Figure 4. illustrates how $\mu$ values depend on $N$ and $H_0$ for a given characteristic time for cloud recovery time ($\tau$ = *20 min).* $\mu_{lim}$ is marked by the magenta contour. The $\mu < \mu_{lim}$ subspace, to the right of the contour, represents $\mu$ values where $Re\{\beta\}$ values can have positive values and so the system can have unstable fixed points (depending on the $D$ values). For larger $\tau$ values, the $\mu_{lim}$ contour shifts to the left, expanding the relative area over which the system can be unstable.

However, within the possibly unstable subspace in which $\mu < \mu_{lim}$ the system's response to changes in $\tau$ were shown to exhibit non-monotonic behavior, namely an increase in $\tau$ will shift the solution from a stable to an unstable, and back to a stable fixed point state. Both $D*$ and $\mu$ are linear in $1/\tau$ and therefore varying $\tau$ corresponds to moving along a straight, inclined line via the origin in the $\mu$ vs. $D*$ parameter space (dotted magenta line in Fig. 1 'c'). The slope of such lines depends on the value of $\sqrt{N}/H_0$. Smaller $\sqrt{N}/H_0$ values imply a steeper slope and therefore a longer path in the non-stable fixed point regime, again showing that smaller aerosol concentrations reduce the size of the parameter regime in which the system is in a steady state.

We acknowledge that such a simplified description of the interplay between cloud and rain cannot capture the full complexity of these interactions in natural systems. Moreover, some of the physical processes that are important in observed clouds are not included here (radiation, for example). Nevertheless, in the spirit of simple dynamical systems analogues to complex systems, the analytical and numerical analyses shown here offer new perspectives on the role and interplay of the main physical parameters. The work demonstrates how changes in these parameters can shift the system between different regimes within the solution space. The existence of such regimes could be explored in the future with more detailed numerical models.

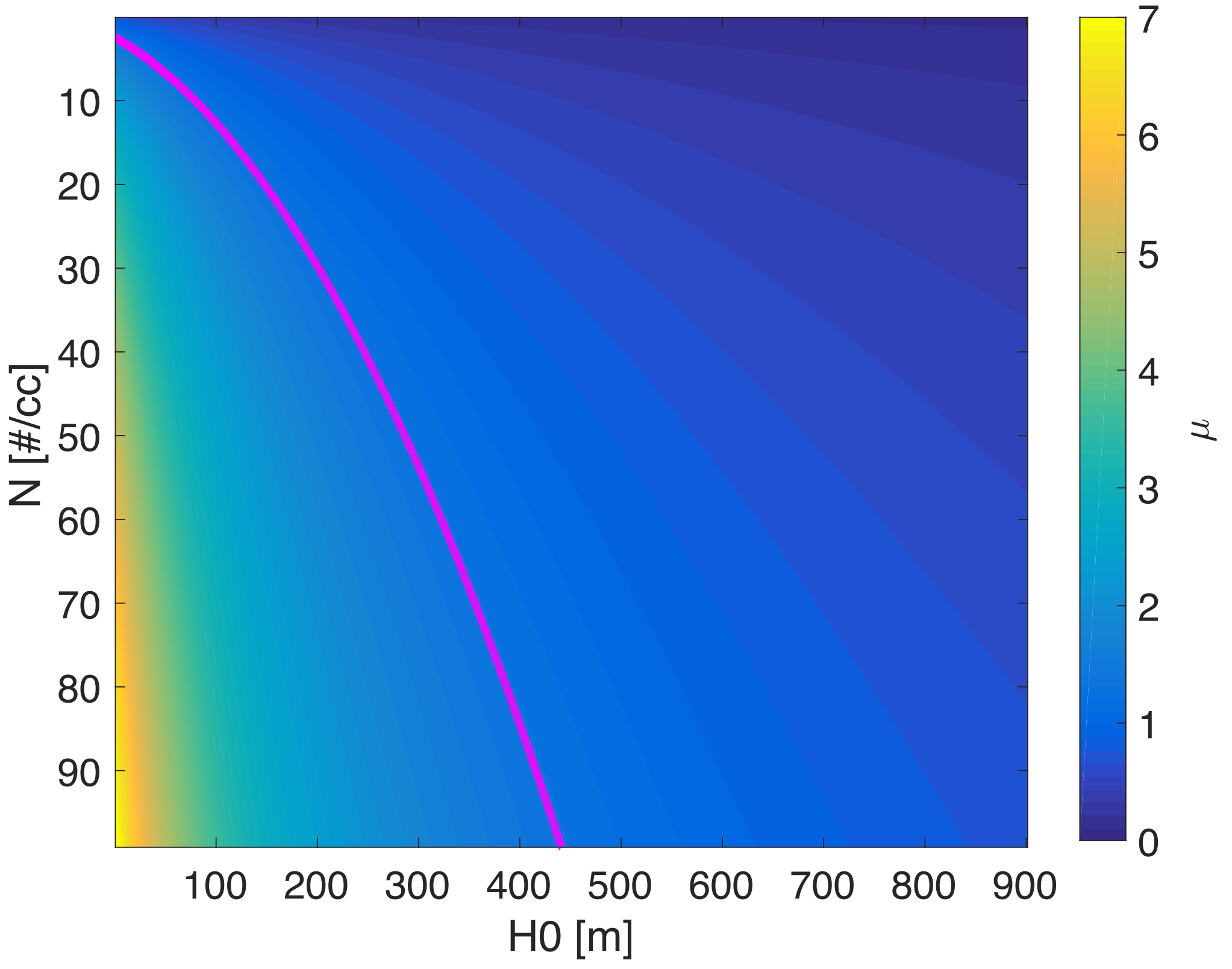

*Figure 4. μ values as a function of $H_0$ and N. The magenta contour marks the $\mu_{lim} = 4/3$ line. The subspace for which $\mu < \mu_{lim}$ (right of the magenta contour) is where unstable fixed points can be found. These are associated with high $H_0$ and small N, i.e., the regime where rain is more likely.*

Acknowledgments

The research leading to these results received funding from the European Research Council (ERC) under the European Union's Seventh Framework Programme (FP7/2007-2013)/ERC Grant agreement no. 306965 (CAPRI). ET is funded by the NSF Physical Oceanography program, grant OCE-1535800, and thanks the Weizmann Institute for its hospitality during parts of this work.

## Appendix A

Lemma 1: An extended view of the solution form of the Lambert $W$ function for any real number. The Lambert $W$ function maps $ze^z \rightarrow z$. If

$$W(\eta) = z, \forall \eta \in \mathbb{R} \; and \; z \in \mathbb{Z}, \tag{L1}$$

then

$$ze^z = \eta. \tag{L2}$$

Expressing $z = x + iy$ and requiring that $Im(ze^z) = 0$ implies the following link

$$x = \frac{-y}{\tan y}. \tag{L3}$$

Inserting Eq. L3 in L2 yields

$$\frac{1}{\cos y} xe^x = \eta, \tag{L4}$$

Therefore Eq. L1 becomes

$$W\left(\frac{1}{\cos y} xe^x\right) = x + iy \tag{L5}$$

Eq. L4 suggests an extended view on the Lambert W function adding a factor of $\frac{1}{\cos y}$ to the standard $xe^x$ kernel of the function. Moreover $\left|\frac{1}{\cos y}\right| \geq 1, \forall y$. For the cases that $\eta \geq \frac{-1}{e}$, a real solution for $W(\eta)$ exist for which y=0 and $\frac{1}{\cos y} = 1$, such that the solution collapse to the standard form of $W(\eta) = x \; or \; \eta = xe^x$. In the general case however, the fact that $\left|\frac{1}{\cos y}\right| \geq 1$ implies that for $\eta > \frac{1}{e}$ , $W(\eta) \geq Re\{W(-\eta)\}$. Such representation for the cases of real arguments and the derived results may have implications in other cases where solutions of the W function are involved. We note that the solution presented in Eq. 5 is not unique and that there are other W function's branches that yields x,y pairs that can satisfy the equation. The above theorem is applicable to all of these x,y pairs.